\documentstyle[preprint,aps]{revtex}
\begin{document}
\draft

\title{Oscillatory Behavior of Critical Amplitudes of the Gaussian Model
on a Hierarchical Structure}

\author{Milan Kne\v zevi\'c}
\address{Faculty of Physics, University of
Belgrade, P.O.Box 368,\\ 11001 Belgrade, Serbia, Yugoslavia}
\author{Dragica Kne\v zevi\' c}
\address{Faculty of Natural Sciences and
Mathematics, University of Kragujevac,\\ 34000 Kragujevac, Serbia, Yugoslavia}
\date{May 18, 1999}

\maketitle
\begin{abstract}

We studied oscillatory behavior of critical amplitudes for the Gaussian model on a hierarchical structure presented by a modified Sierpinski gasket lattice. This model is known to display non-standard critical behavior on the lattice under study. The leading singular behavior of the correlation length $\xi$ near the critical coupling $K=K_c$ is modulated by a function which is periodic in $\ln\vert\ln(K_c-K)\vert$. We have also shown that the common finite-size scaling hypothesis, according to which for a finite system at criticality $\xi$ should be of the order of the size of system, is not applicable in this case. As a consequence of this, the exact form of the leading singular behavior of $\xi$ differs  from the one described earlier (which was based on the finite-size scaling assumption).  

\end{abstract}

\pacs{64.60.Ak, 05.50.+q, 05.40.Fb}

\newpage
\narrowtext
\section{Introduction}
\label{sec:intro}

In the near past a considerable research activity has been devoted to the studies of recursion relations which have a singular structure near the pertinent fixed points [1-6]. It was found that, under certain conditions, these singularities can lead to an unusual critical behavior of relevant  physical quantities. In particular, it has been shown that the mean end-to-end distance $R_{_N}$ of a simple ideal polymer chain  on some hierarchical structures can grow more slowly than any power of its length $N$. This effect has been termed localization, and it has been attributed to an entropic trapping of the polymer chain \cite{Mar}: In order to maximize the entropy, it is advantageous for a chain to visit the lattice sites of the highest coordination number preferentially. These sites act, therefore, as entropic traps preventing the swelling of the chain.

It is well known that statistics of an ideal polymer chain on lattices can 
be captured by means of a suitable Gaussian model. Using this connection a number of interesting results for the polymer model have been derived by studying the singular structure of associated recursion relations for the  Gaussian model. Let us call here just one example: It has been argued, by using finite-size scaling arguments and an analytical study of a pertinent mapping, that the mean end-to-end distance of an ideal chain on the modified  Sierpinski gasket (see Fig. 1) follows the logarithmic asymptotic law: $R_{_N}\sim \ln^\Phi N$, with $\Phi=\ln 3/\ln 2$.  As it has been emphasized \cite{KK}, however, it was very difficult to check this result numerically, due to the oscillations in the values of the Gaussian correlation length $\xi$. In fact, these oscillations are so pronounced that they mask even leading singular behavior of $\xi$.

This type of corrections to the leading asymptotic behavior near criticality were studied quite early \cite{NvL}, within the framework of the renormalization group approach. Recently, they were observed in many systems displaying the power law singularities, and they were related to the concept of discrete scale invariance \cite{Sor}. This motivated us to reconsider here the Gaussian model on a modified SG lattice. In this case the hierarchical structure of the lattice leads to the oscillations of critical amplitudes of various quantities close to criticality. If one presents them on an appropriate scale, it turns out that they become regularly spaced.

We have found that the above mentioned oscillations are universal, and that they can be described in terms of some simple functions which are periodic in the variable 
$\ln\vert\ln(\delta K)\vert$ (with $\delta K=K_c-K$ being the distance from the critical point). This is in contrast to the critical behavior of usual systems (displaying the power law singularities), in which case corresponding variable has the form $\ln(\delta K)$. What is perhaps even more interesting, we have found that our numerical values of the correlation length do not fit the form $\xi_0\ln^\Phi(\delta K)$, with the above reported value $\Phi=\ln 3/\ln 2$. We have shown instead, using an asymptotic matching, that one has to take $\Phi=\ln(3/2)/\ln 2$  in order to have a proper agreement between analytical and numerical results. Consequently, the common finite-size scaling arguments, which were used in the previous studies of this model \cite{Mar,GMN,KK}, are not applicable in this case.

In Sec. II we present our model, recall some previously obtained results, and examine oscillatory behavior of some critical amplitudes. Our conclusions are given in Sec III.

\section{THE MODEL AND ITS ANALYSIS}
\label{sec:model}
We consider here the usual zero-field Gaussian model on a hierarchical structure which is presented by a modified Sierpinski gasket (SG) lattice of base $b=3$ (Fig. 1). Partition function of this model has a simple form,
\begin{equation}
{\cal Z}(K)=\sum_{-\infty}^{\infty}{dS_1\dots dS_N\,\exp
[-\frac{1}{2}\sum_i{S_i^2}+K\sum_{<ij>}{S_iS_j}]},
\end{equation} 
where $S_i$ is the continuous spin variable at site $i$, 
$K=J/k_BT$ where $T$ is temperature and $J$ represents interaction between each nearest-neighbor pair of Gaussian spins, while $<ij>$ denotes the summation over all such pairs. As we have shown in a previous paper \cite{KK}, the $r$th order partition function can be expressed in terms of three parameters $A^{(r)}$, $B^{(r)}$, and $D^{(r)}$ which obey a set of recursion relations and suitable initial conditions. 
These relations are somewhat cumbersome (see Eqs. (34)-(35) of \cite{KK}), and  we will not repeat them here. Let us note, nevertheless, that they have a singular structure near the relevant fixed point, which does not allow us to make a common fixed-point analysis. As detailed in \cite{KK}, for the critical value $K_c=0.227\,148\dots$ of the interaction strength $K$, all successive iterations of $A^{(r)}$ and $B^{(r)}$ lie on an invariant line, which starts at the point $(A^{(0)}=0,B^{(0)}=K_c)$ and ends at the fixed point $(A^*=1/6,B^*=0)$. An asymptotic equation of this line, which is valid near the fixed point, has been found perturbativelly \cite{KK}. One can show that along this line ($\delta K=0$) parameter $B$ renormalizes according to the law
\begin{equation}
B'=2\sqrt{3}B^2+(18+12\sqrt{3})B^3-\frac{21}{4}\sqrt{3}B^4+\dots,
\end{equation}
while away from it ($\delta K>0$) this parameter follows the law: 
$B'\sim B^3$. It is evident that the solution $B_r\sim \mu^{3^r}$ (where $\mu=\mu(\delta K)<1$) well fits in with the latter condition. On the other hand, an asymptotic solution of (2) can be expressed as a power series in $\kappa^{2^r}$,
\begin{equation}
B^{(r)}=\frac{\sqrt{3}}{6}\kappa^{2^r}-\frac{2+\sqrt{3}}{8}\kappa^{2\cdot 2^r}+
\frac{312+193\sqrt{3}}{384}\kappa^{3\cdot 2^r}+\dots,
\end{equation}
where $0<\kappa<1$ is a constant that can be determined numerically. Although this solution has been derived for $\delta K=0$, it holds also for a finite but very small value of $\delta K$ ($0<\delta K\ll 1$), provided $r\lesssim r_0$, where $r_0\gg 1$ is the number of iterations one can make along the invariant line before going away from it. These two regimes are separated by a rather narrow crossover region, which makes it possible to apply the asymptotic matching: 
\begin{equation}
\mu^{3^{r_0}}=\exp[\ln(\mu)\, 3^{r_0}]\sim \kappa^{2^{r_0}}=\exp[\ln (\kappa)\,2^{r_0}],
\end{equation}
i.e.
\begin{equation}
\xi\sim\left(\frac{3}{2}\right)^{r_0},
\end{equation}
where $\xi(\delta K)=-1/\ln [\mu(\delta K)]$ stands for the correlation length of the model \cite{Kor}. 
This finding is in contrast to the usual finite-size scaling expectation, according to which the correlation length of a finite system at criticality should be of the order of system's size ($\xi\sim 3^{r_0}$). 

The number of iteration $r_0$ along the invariant line depends on the value of $\delta K$ and can be estimated from the  obvious relation:
$B^{(r_0)}(\delta K)\approx B^{(r_0)}(0)+\left.\frac{dB^{(r_0)}}{dK} \right\vert_{\delta K=0}\delta K$.
Indeed, taking into account (3) and relation  $dB^{(r_0)}/dK\sim 2^{r_0}$  (see \cite{KK}), we find: 
\begin{equation}
\kappa^{2^{r_0}}\sim 2^{r_0}\delta K,\quad\text{or}\quad 2^{r_0}\ln(\kappa)\sim
\ln(\delta K).
\end{equation}
This, together with (6), leads to a logarithmic singular behavior of $\xi$
\begin{equation}
\xi\sim \vert\ln(\delta K)\vert^\Phi,\quad\text{with}\quad \Phi=\frac{\ln(3/2)}{\ln 2}.
\end{equation}
 
This differs (in the value of $\Phi$) from the earlier reported results which  have been derived by using the finite-size scaling assumption: $\xi\sim 3^{r_0}$.  This difference is caused by the peculiar behavior of the parameter $B$ and correlation function -- in our case they decrease exponentially to zero even at criticality (see (3) and note \cite{Kor}).

In order to provide some further insight into the critical behavior of the model, we will also examine it numerically. Thus, using a huge precision, we have been able to come very close to the critical point ($\delta K/K_c<10^{-3000}$). This allows us to calculate the correlation function and associated correlation length in a wide region (on a logarithmic scale) around the critical point \cite{Numer}. Our results are presented in 
Fig. 2(a), where the scaled correlation length (critical amplitude $\xi_0$) $\xi_0=\xi \vert\ln(\delta K)\vert^{-\Phi}$ is displayed as a function of $\ln\vert\ln(\delta K)\vert$. The overall behavior of $\xi_0$ is highly sensitive to the precise value of $\Phi$. For example, average value  $\overline\xi_0=(\xi_{0,max}+\xi_{0,min})/2$ of $\xi_0$ appears to be a constant for sufficiently small values of $\delta K$, with the above quoted value of $\Phi$ (see Fig. 2(a)), while $\overline\xi_0$ becomes unstable under a small change of $\Phi$. This provides a good criterion for a numerical calculation of $\Phi$. Indeed, in this way we have been able to determine $\Phi$ with four correct digits, and a further improvement depends on the possibility to approach the fixed point still more closely. This should be contrasted with the straightforward procedure \cite{KK}: A plot of $\ln\xi$ versus $\ln\vert\ln(\delta K)\vert$ leads to 
numerical estimates of $\Phi$ which oscillate with large amplitudes around the exact value, independent on the distance $\delta K$ from the critical point.

It seems that $\xi_0$ represents a simple periodic function of $\ln\vert\ln(\delta K)\vert$. Period of this function, estimated numerically, is found to be in excellent agreement with the theoretical value $\tau=\ln 2$. Perhaps the simplest way to understand this is to adopt the following point of view: One can regard (7) as a 'pure' power law, with the scaling variable $\vert\ln(\delta K)\vert$ (rather than $\delta K$) and a 'critical exponent' $\Phi$. In the same spirit, one can interpret the relation (6) as $\lambda^{r_0}\sim\vert\ln(\delta K)\vert$, with $\lambda=2$ playing the role of a 'thermal' eigenvalue, while relation (5) provides an 'effective' spacial scaling ratio $3/2$. It is clear then, from standard theory of log-periodic corrections to the power law scaling \cite{Sor}, that critical amplitude $\xi_0$ should be a periodic function in $\ln\vert\ln(\delta K)\vert$, with  period $\tau=\ln\lambda=\ln 2$.

We have also analysed the critical behavior of the first derivative of the free energy density with respect to $K$ (internal energy $E$). Using the approach described in \cite{KK}, we have found that this quantity exhibits an interesting confluent singularity,
\begin{equation}
E\sim\frac{1}{\delta K}\vert\ln(\delta K)\vert^{-\Psi},\quad\text{with}\quad
\Psi=\frac{\ln 6}{\ln 2},
\end{equation}
which corresponds to a first order phase transition. 
As in the case of the correlation length, we have studied the internal energy numerically. Our results are displayed in Fig. 2(b), where we presented the scaled energy (i.e. critical amplitude $E_0$) $E_0=\delta K\vert\ln(\delta K)\vert^{\Psi}E$ as a function of $\ln\vert\ln(\delta K)\vert$. It is evident that $E_0$ represents a simple periodic function of  $\ln\vert\ln(\delta K)\vert$, while its period is in agreement with the above quoted theoretical value ($\tau=\ln 2$). At the same time this analysis provides a good numerical check of the form (8) of the energy leading singularity.   

\section{CONCLUSION}
In this paper we have studied critical behavior of the Gaussian model on a modified SG. We have shown that both correlation length and energy critical amplitudes exhibit very pronounced oscillations near the critical coupling $K_c$. These oscillations can be described in terms of some simple functions which are periodic in $\ln\vert\ln(\delta K)\vert$. Period $\tau$ of these functions is found to be determined by a universal quantity which governs  critical behavior of the model. Knowledge of these functions is very useful because it provides  more precise description of the critical behavior of quantities under consideration. 

Having determined the basic properties of these functions, we have been able to make a precise numerical check of the exact form of the leading singular behavior of $\xi$ and $E$. In particular, this allowed us to notice an inaccuracy in a previously described leading asymptotic form of $\xi$ near criticality \cite{Mar,GMN,KK}. It seems that this discrepancy stems from the inapplicability of the standard finite-size scaling assumption in this model. Indeed, using a simple technique, not relying on this assumption (an asymptotic matching), we derive somewhat different singular behavior of $\xi$ (see (7)), which turns out to be in excellent agreement with the acquired numerical findings (Fig. 2). This example points out that an 'ad hoc' use of finite-size scaling assumptions could be questionable sometimes, and that one has to use them with caution in general case.

\vfill
\eject

\vfill
\eject

\begin{figure}
\caption{First two stages in the iterative construction of the modified $b=3$ Sierpinski gasket.}
\label{fig1}
\end{figure}
\begin{figure}
\caption{(a) The correlation length critical amplitude $\xi_0$, in units of  lattice constant, as a function of $\ln\vert\ln(\delta K)\vert$ ($\xi_0=\xi \vert\ln(\delta K)\vert^{-\Phi}$,  $\Phi=\ln(3/2)/\ln2$). (b) The energy critical amplitude $E_0$, in units of the interaction strength $J$, as a function of the same variable ($E_0=\delta K\vert\ln(\delta K)\vert^{\Psi}E$, $\Psi=\ln 6/\ln 2$). These functions are computed from exact representations of related quantities by using a huge numerical precision [10]. Note that the amplitudes of these oscillations are so large that they, in fact, mask the leading asymptotic behavior of $\xi$ and $E$.}     
\label{fig2}
\end{figure}


\begin{references}
\bibitem{Kor} One can show  that two-point correlation function $G^{(r)}$ (see Eq. (36) of [5]) has the same asymptotic behavior as parameter $B^{(r)}$.
\bibitem{Numer} All our calculations were done in Mathematica. In particular, we have used facilities provided by this package in doing with arbitrary-precision numbers. In this way we have been able to handle a highly unstable arithmetic which appears when we iterate our recursion relations.
Taking into account that the accuracy of our output results degrades progressively when we approach the fixed point more and more closely, we had to use a formidable precision. For example, in order to determine $K_c$ with $3000$ correct digits we set up the precision of our input data to $6000$ digits.
\end{references}
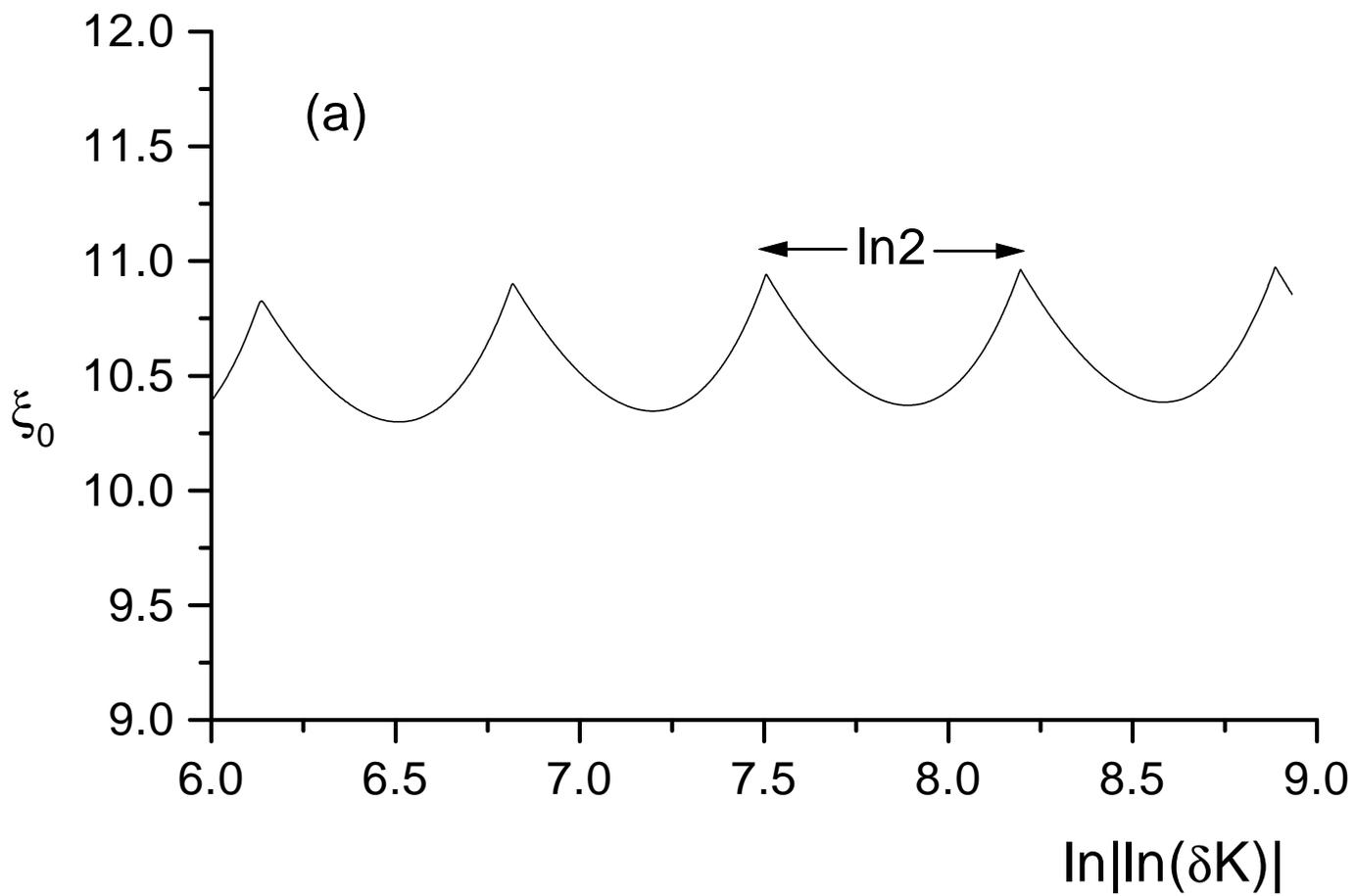
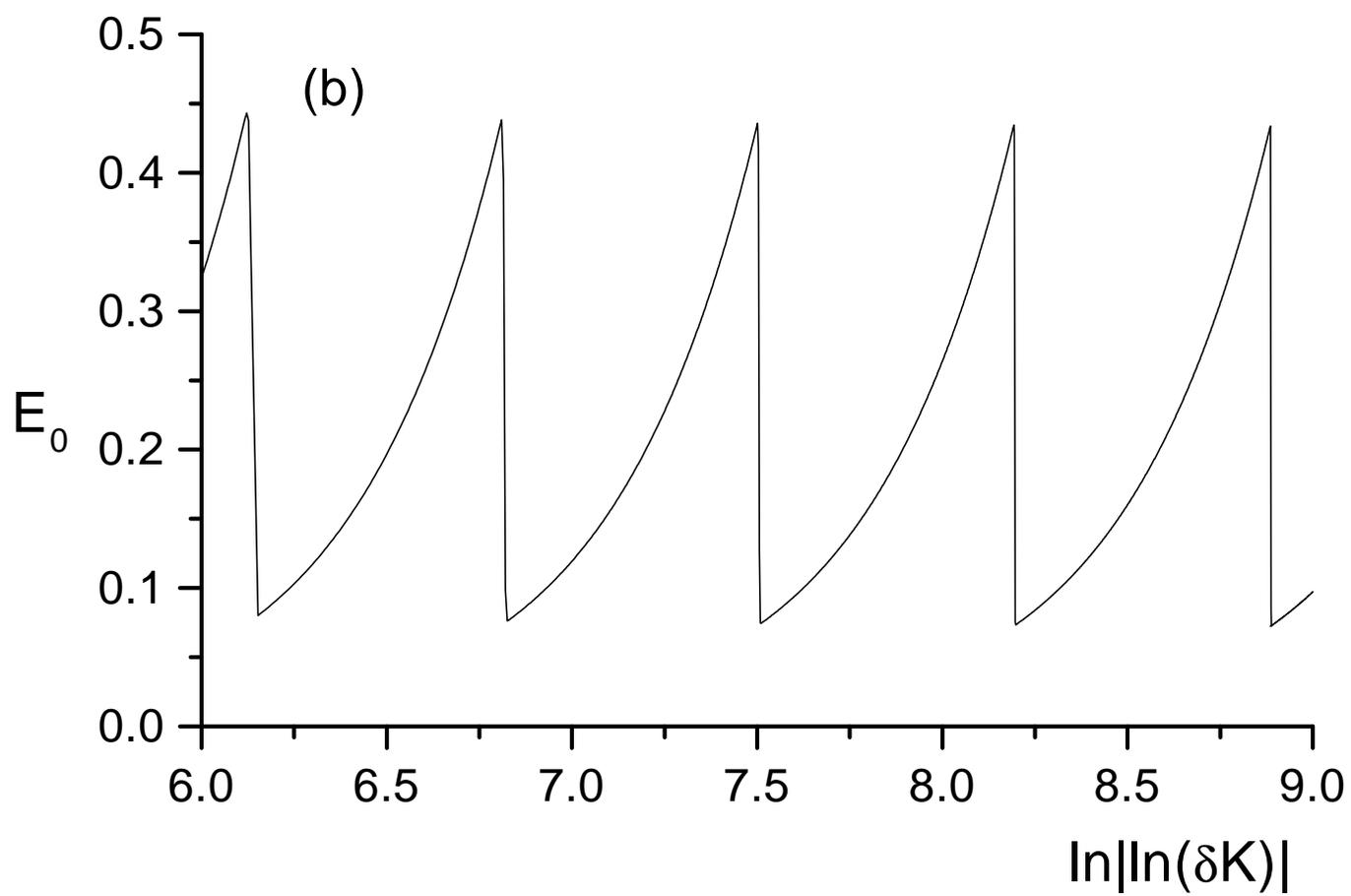
\end{document}